# Mechanical Design of the PIP-II ORBUMP Pulsed Dipole Magnet

D. Karas*, K. Badgley, Z. Chen, V. Chernenok, M. Davidson, D. Harding, D. Johnson, V. Kashikhin, W. Robotham, T. Strauss, B. Szabo, J. Vander Meulen – Fermi National Accelerator Laboratory

*Abstract*—The Proton Improvement Plan II (PIP-II) project is a vital upgrade to the Fermilab accelerator complex. The magnet pulse rate of the PIP-II Injection system requires an increase from the current rate of 15 Hz to 20 Hz as well as a roughly 30% increase in the magnetic field of the new Orbital Bump (ORBUMP) pulsed dipole magnets in the Booster. The ORBUMP magnet mechanical design is presented in this paper. The ORBUMP magnet is secured in a vacuum box and the core is made up of 0.127 mm thick, low carbon steel laminations with a C-5 inorganic magnesium phosphate coating. The core is clamped using external tie bars welded to the core end plates. ANSYS Finite Element Analysis (FEA) was used to evaluate the clamping design to minimize the deflection of the core post welding of the tie bars. The water-cooled, single turn coil, which shapes the magnetic field by acting as the pole tips, is critical for the integrated field homogeneity. The coil manufacturing tolerances and fabrication techniques were evaluated to ensure the magnetic properties of the magnet could be obtained. The coil is electrically isolated from the core using virgin Polyether ether ketone (PEEK) insulating material in the gap. An investigation into the high voltage performance of the virgin PEEK insulator was conducted via partial discharge testing using a 1:1 scale sample.

## I. INTRODUCTION

THE Proton Improvement Plan II (PIP-II) is a major upgrade to the Fermi National Accelerator Laboratory (Fermilab) accelerator complex. The enhancement will power the world's most intense, high-energy neutrino beam on its journey from Fermilab in Illinois to the Deep Underground Neutrino Experiment (DUNE) in South Dakota [1]. Through the study of neutrinos, Scientists at DUNE aim to answer some of the most fundamental questions about our universe. In addition, PIP-II will enable a diverse and long-term research program, delivering scientific breakthroughs over the next several decades.

The Fermilab Booster injection system was originally designed in the early 1990s with a 3 Hz magnet pulse rate. It was eventually increased to 15 Hz in 2005 in order to meet the intensity and repetition rate requirements of Fermilab's high energy physics program [2] [3]. The PIP-II neutrino beam requires an additional 33.3% increase in the Booster magnets pulse rate to 20 Hz as well a roughly 30% increase of the magnetic field of the new Orbital Bump (ORBUMP) magnets [4]. The beam optics analysis establishes the magnetic parameters outlined in Table 1 and results in a dogleg-shaped injected beam orbit.

TABLE I
ORBUMP MAGNETIC PARAMETERS [4]

| Parameters | Units | Values |
|---|---|---|
| Proton Beam Energy | MeV | 800 |
| Pulse Repetition Rate | Hz | 20 |
| Pulsed Field Ramp Up/Down | ms | 0.2 ± 0.04 |
| Magnetic Field Pulsed Plateau | ms | 0.6 |
| Maximum Integrated Field | T-m | 0.35 |
| Field Homogeneity | % | 1.0 |
| Good Field Area Width | mm | 250 |
| Peak Current | kA | ≤20 |
| Gap | mm | 20 |
| Laminated Core Length | mm | 850 |
| Dimensions (L x W x H) | m | 1.03 x 0.46 x 1.06 |

## II. MECHANICAL DESIGN

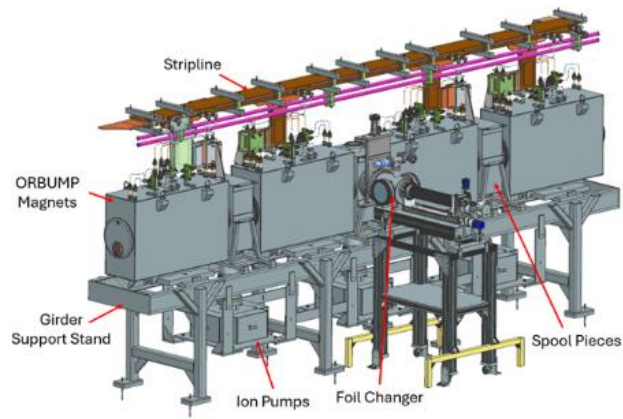

**Fig. 1.** PIP-II injection girder.

The beam orbit is created by four identical vertical bending, single turn ORBUMP magnets connected in series by a power distribution stripline [4]. Fig. 1 illustrates the ORBUMP Booster injection girder which supports the four ORBUMP magnets, spool pieces, ion pumps, and interfaces with the foil changer which strips the beam of H- ions [5]. The water cooled, window-frame ORBUMP magnet is contained inside 316L stainless-steel vacuum boxes as shown in Fig. 2.

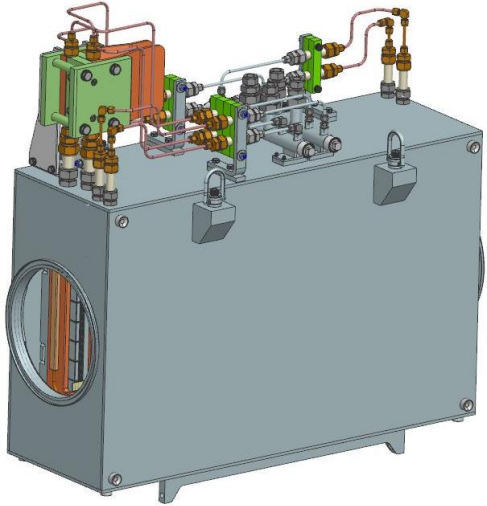

**Fig. 2.** PIP-II ORBUMP magnet.

*A. Core Design and Clamping*

In order to meet the magnetic field and pulse ramp requirements as well as reduce eddy currents, the ORBUMP magnet core is comprised of 0.127 mm thick, low carbon steel laminations with a C-5 inorganic magnesium phosphate coating [4]. The laminations of each half core shall be stacked and clamped together using four external 316L stainless-steel tie bars spanning the length of the core (Fig. 3). The 15.875 mm H x 38.1 mm W tie bars are fillet welded to the 316L stainless-steel end plates and are electrically isolated from the laminations using virgin Polyether ether ketone (PEEK). Each tie bar includes equally sized and spaced cutouts designed to mitigate virtual leaks within the vacuum box.

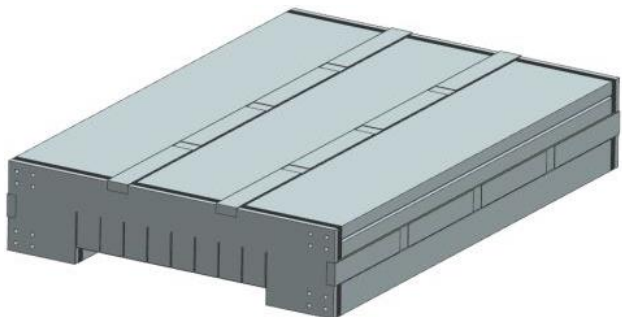

**Fig. 3.** Assembled ORBUMP half core with welded tie bars. The core laminations are shown as a solid body.

An investigation into the deflection of the core post stacking was conducted in order to minimize the deflection to <0.127 mm which is within typical machinable tolerances. The initial design included provisions for clamping the half cores using internal tie rods. However, ANSYS Finite Element Analysis (FEA) revealed a large deflection and subsequent fanning of the laminations across the half core mating plane [6].

In the initial design, three electrically isolated M14 x 2, 18-8 stainless-steel tie rods were equally spaced along the outer perimeter of each half core and were fastened to the end plates. To simplify the model in ANSYS, the fastening hardware and insulating material were removed while the laminated core was represented by eighty-five 10 mm sections. Linear elastic material properties were assigned to the components using ANSYS materials library. Frictional contacts with a conservative static coefficient of friction of 0.2 were used throughout the model. This was chosen 82 kN of force was applied to the interface between the tie rod washers and the end plates on one side which simulates the reaction force from stacking the laminations. The same surfaces were used as fixed supports on the opposite side. The maximum directional deflection in the X direction with respect to the core parting plane using tie rods was evaluated to be approximately 0.156 mm as shown in Fig. 4.

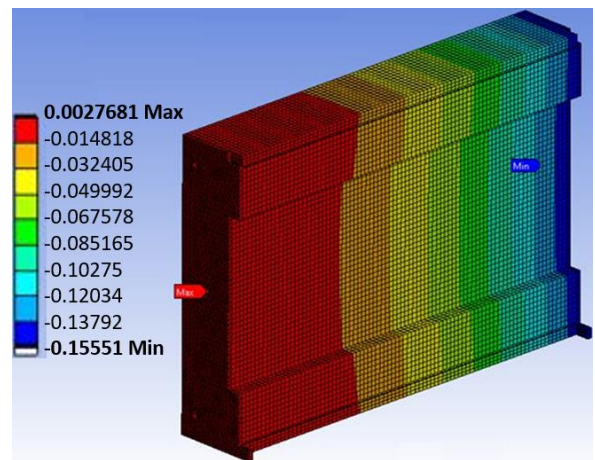

**Fig. 4.** X directional deflection with respect to the core parting plane with initial tie rod clamping design (True Scale).

The large deflection was deemed insufficient by beam physicists because it would disrupt the field homogeneity and introduce strong higher-order multipole fields. The clamping design was optimized by replacing the tie rods with tie bars to minimize deflection to within the allowable tolerances. For this investigation, the laminated core was once again represented by eighty-five 10 mm sections and frictional contacts with a conservative static coefficient of friction of 0.2 were used throughout the model. The tie bars and insulating material were removed from the simulation in an effort to simplify the simulation. Linear elastic material properties were once again assigned to the components using ANSYS materials library. 82 kN of force was applied to the interface between the tie bars and the end plates on one side which simulates the reaction force from stacking the laminations. The same surfaces were used as fixed supports on the opposite side. The maximum directional deflection in the X direction with respect to the core parting plane was evaluated to be approximately 0.098 mm (Fig. 5). This is within the allowable tolerances stated previously.

Fabrication of a prototype core is underway at Fermilab. Once complete, it will undergo full dimensional inspection to compare against the ANSYS FEA results. The simplified FEA model used 85 sections to represent the core consisting of 6693 laminations. This was necessary to decrease the number of contacts to save computational power but resulted in stiffer simulation results. The conservative static coefficient of friction 0.2 was chosen to run the simulations as the ANSYS simulation could not solve higher frictional values. The results from the dimensional inspection of the prototype core will ultimately verify the ANSYS FEA reliability.

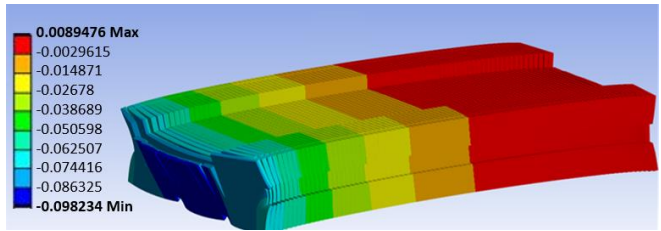

**Fig. 5.** X directional deflection with respect to the core parting plane with tie bar clamping design (Auto Scale).

*B. Coil Design and Fabrication*

The ORBUMP magnet coil is a water cooled, single turn coil as shown in Fig. 6 and made out of oxygen free copper. The coil consists of two conductor bars, two jumper plates, leads, positional locating keys as well as two water cooling channels. The jumper plates contain clamps to retain the water channels while ensuring adequate cooling to those components. The water channels are also embedded into the conductor bars as illustrated in Fig. 7. The coil is electrically isolated from the core in the gap using virgin PEEK insulating material. Since four ORBUMP magnets are powered in series by the power distribution stripline, the average heat load of the entire coil at the nominal operating current was found to be 1.54 kW per magnet during OPERA3D simulations [4]. Cooling is provided by forced flow, low conductivity water with a nominal inlet temperature of 32°C through the 6.35 mm outer diameter x 0.66 mm wall water channels. The heat transfer coefficients were determined via water flow characteristics and thermal simulations using ANSYS indicate the peak coil temperature will be roughly 52°C.

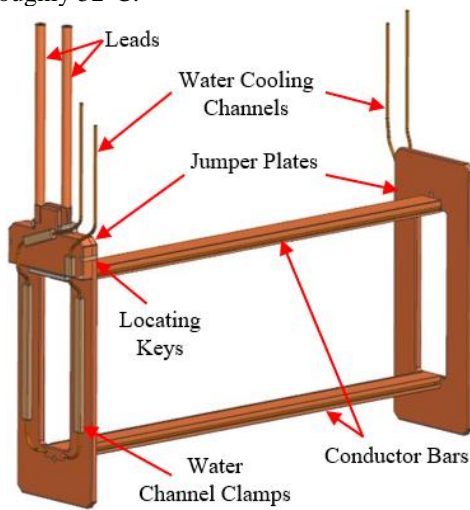

**Fig. 6.** ORBUMP Coil.

The full coil, including the water channels, shall be vacuum oven brazed to ensure thermal expansion and subsequent contraction is consistent throughout all subcomponents. Vacuum oven brazing of the full coil requires a brazing fixture to hold all positional tolerances of the coil subcomponents. The water channels shall be tack brazed to the jumper plates and conductor bars prior to oven brazing to reduce the complexity of the brazing fixture

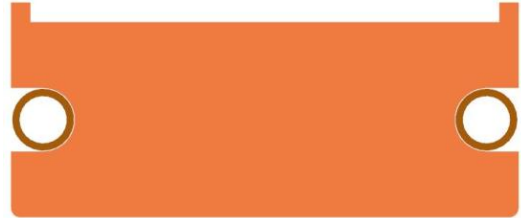

**Fig. 7.** Conductor bar cross section.

The conductor bars shape the magnetic field. The field homogeneity requirements were met by including 2 mm x 2 mm shims on the top surface of the conductor bars (Fig. 7) [4]. Properly tolerancing the flatness and parallelism the brazed coil conductor bars determine the field homogeneity of the magnet during operation. Due to the thermal characteristics of the coil during brazing, post braze processing is imperative to ensure the profile tolerances and field homogeneity requirements are met. Wire EDM with proper fixturing is the preferred post-braze processing method in order to hold all tolerances.

## II. PARTIAL DISCHARGE TESTING

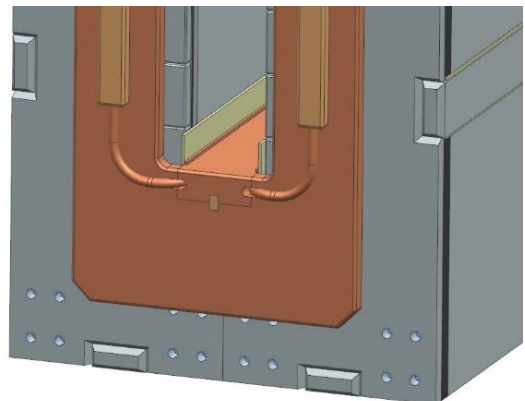

**Fig. 8.** ORBUMP coil with virgin PEEK insulator installed in ORBUMP magnet gap.

As stated previously, the coil is electrically insulated from the core using virgin PEEK insulating material as shown in Fig. 8. An investigation into the high voltage performance of the virgin PEEK was conducted. A 1:1 scale section of the copper conductor bar, virgin PEEK insulating material, and steel U-channel representing the core were manufactured for testing. Testing was conducted using a 60 Hz AC waveform generated by Hipotronics HDA5 capable of 5 kV rms. An oscilloscope was used to measure response from the HDA5. Tests were conducted both in air and under high vacuum. The test results examine the lowest voltage where partial discharge is observed continually (i.e., the inception voltage). Air tests were completed to understand the voltage limitations of the magnetic field measurements of each individual magnet which is to be performed in air. Vacuum tests were performed to understand the voltage limitations of the magnets under normal operation.

A baseline test in air was performed by inserting rounded, conductive material mounted to the high voltage electrode into the virgin PEEK insulator over a grounded copper plane (Fig. 9). This was conducted to evaluate the highest voltage the insulating material can achieve at the given thickness and was compared against the inception voltage of the conductor sample. This test achieved 3.5 kV rms without inception, except under prolonged exposure.

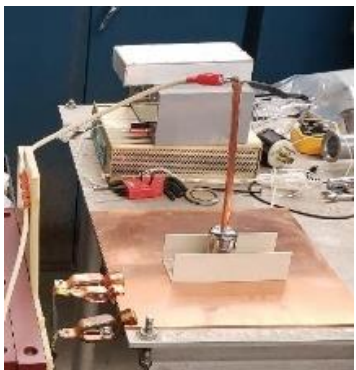

**Fig. 9.** Baseline partial discharge test

The next several tests were carried out by placing the conductor sample within the virgin PEEK insulator at various distances from the edge over a grounded copper plane. Testing was also conducted by placing the conductor and virgin PEEK insulator within the steel U-channel to understand the voltage limitations of the magnetic field measurements as stated previously. Fig. 10 depicts the conductor located in the center of the insulator. The inception voltage was reduced to 2.9-3.0 kV rms and a low of 2.3 kV rms when the conductor was placed at the center and flush with the edge of the insulator respectively. The reduction in inception voltage can be attributed to the air layer between the virgin PEEK insulator and steel U-channel as well as the overall geometry of the conductor sample. The sample is a 'slice' of the overall conductor bar design which introduces a number of edges (i.e., right angles) and vertices leading to electric field concentrations in these areas.

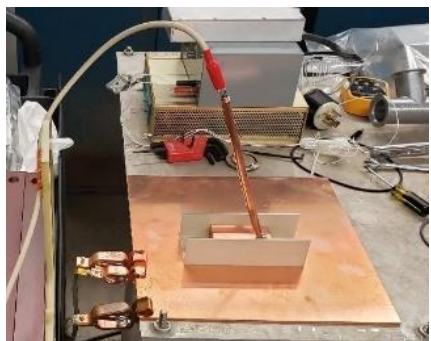

**Fig. 10.** Conductor placed at center of virgin PEEK insulator.

To qualify the voltage limitations of the magnets under normal operations, the conductor bar was inserted into the virgin PEEK insulator and steel U-channel for testing within a vacuum chamber. The sample (Fig. 11) was cleaned with Ethyl alcohol and lint free vacuum wipes prior to insertion into the vacuum chamber (Fig. 12). Testing was sustained for over 1 minute under vacuum levels ranging from $10^{-6}$ to $10^{-3}$ Torr. No partial discharge was observed within these vacuum ranges, which minimizes concerns for partial discharge within the magnet while operated in sufficient vacuum.

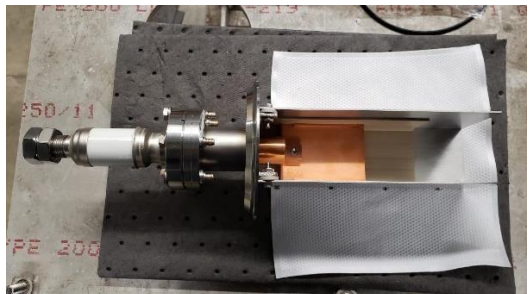

**Fig. 11.** Sample for partial discharge testing under vacuum

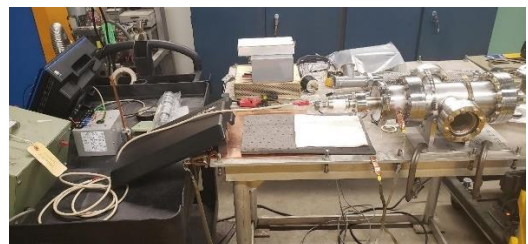

**Fig. 12.** Partial discharge testing within vacuum chamber

These tests can only indicate the inception voltage. No information about the 'amount' of partial discharge was obtained. The high voltage waveform the coil will see during normal operation is a 2.7 kV rms impulse. The peak voltage at 60 Hz shows a withstanding rating >2.5 times the peak voltage the magnet will see while under sufficient vacuum during normal operation

## V. CONCLUSION

The proposed designs and fabrication techniques, as verified through ANSYS FEA and partial discharge testing, have been shown meet the magnet requirements. The next critical steps include fabricating a prototype core and coil to compare to the simulations and test results presented in this paper.

ACKNOWLEDGMENT

The authors would like to thank the US Department of Energy Office of Science, as well as the Fermilab teams and management, for their support in completing this work and overall upgrade to the Fermilab accelerator complex.